\documentclass[aps,preprint,superscriptaddress,nofootinbib,preprintnumbers,prd]{revtex4-1}

\usepackage{amsmath,amssymb,amsfonts}
\usepackage{bm, slashed, array,multirow}
\usepackage{xcolor,graphicx,placeins}
\definecolor{hgreen}{rgb}{0,0.3,0}
\definecolor{hred}{rgb}{0.3,0,0}
\definecolor{hblue}{rgb}{0,0,0.3}
\usepackage[colorlinks=true,
            linkcolor=hblue,
            citecolor=hgreen,
            filecolor=hblue,
            urlcolor=hred]{hyperref}

\newcommand{\bkg}{B \to K^*\gamma}
\newcommand{\bbkg}{\bar{B} \to \bar{K}^*\gamma}
\newcommand{\kkp}{K^* \to K^+\pi^-}
\newcommand{\kbkp}{\overline{K}^* \to K^-\pi^+}

\newcommand{\mk}{m_{K^*}}
\newcommand{\Gk}{\Gamma_{K^*}}

\newcommand{\mb}{m_B}
\newcommand{\me}{m_e}

\newcommand{\gpl}{g_{\parallel}}
\newcommand{\gpp}{g_{\perp}}
\newcommand{\gbpl}{\bar{g}_{\parallel}}
\newcommand{\gbpp}{\bar{g}_{\perp}}
\newcommand{\gks}{g_{K^*}}

\newcommand{\mcM}{\mathcal{M}}
\newcommand{\mcN}{\mathcal{N}}
\newcommand{\sh}{\hat{s}}
\newcommand{\bfull}{B \to (K^* \to K^+\pi^-) (\gamma \stackrel{\rm{BH}}{\to} e^+e^-)}
\newcommand{\bbarfull}{\bar{B} \to (\bar{K}^* \to K^-\pi^+) (\gamma \stackrel{\rm{BH}}{\to} e^+e^-)}

\newcommand{\mcA}{\mathcal{A}}
\newcommand{\mcC}{\mathcal{C}}
\newcommand{\mcS}{\mathcal{S}}
\newcommand{\mcT}{\mathcal{T}}
\newcommand{\mcCc}{\mcC_{\textrm{c}}}
\newcommand{\mcSc}{\mcS_{\textrm{c}}}
\newcommand{\mcTc}{\mcT_{\textrm{c}}}
\newcommand{\Nc}{N_{\textrm{c}}}

\newcommand{\asbr}[4]{ \langle #1^#2| #3^#4 \rangle}
\newcommand{\sigbr}[4]{ \langle #2^#1| #3 | #4^#1 \rangle}

\begin{document}

\preprint{FERMILAB-PUB-15-142-T}

\graphicspath{{./figs/}}

\def\Berkeley{Department of Physics, University of California, Berkeley, CA 94720, USA}
\def\LBL{Ernest Orlando Lawrence Berkeley National Laboratory, University of California, Berkeley, CA 94720, USA}
\def\Cincy{Department of Physics, University of Cincinnati, Cincinnati, Ohio 45221, USA}
\def\Fermilab{Theoretical Physics Department, Fermilab, P.O. Box 500, Batavia, IL 60510, USA}

\title{Probing the photon polarization in \texorpdfstring{$B \to K^*\gamma$}{BKG} with conversion}

\author{Fady Bishara}              
\email[E-mail:]{bisharfy@mail.uc.edu}
\affiliation{\Cincy}
\affiliation{\Fermilab}

\author{Dean J. Robinson}
\email[E-mail:]{djrobinson@berkeley.edu}
\affiliation{\Berkeley}
\affiliation{\LBL}

\date{\today}

\begin{abstract}
We re-examine the possibility to measure the photon polarization in $B \to K^*\gamma$ decays, via decays in which the photon subsequently undergoes nuclear conversion to a lepton pair. We obtain compact expressions for the full decay-plus-conversion amplitude. With these results we show that interference between the $B \to (K^*\to K\pi)\gamma$ decay and the $\gamma N \to \ell^+\ell^-N$ conversion permits both the ratio and relative weak phase between the left- and right-handed photon amplitudes to be probed by an angular observable, constructed from the final state dilepton, kaon and pion kinematic configuration. Exploiting this technique will be experimentally challenging. However, we present special kinematic cuts that enhance the statistical power of this technique by an $\mathcal{O}(1)$ factor. We verify this effect and extract pertinent angular kinematic distributions with dedicated numerical simulations. 
\end{abstract}

\maketitle

\section{Introduction}
Measurement of the photon polarization in $b \to s \gamma$ radiative decays has been of long-standing interest. Within the Standard Model (SM), flavor-changing electroweak interactions maximally violate parity, so that one expects the fraction of left-handed photons in $b \to s \gamma$ processes to be order unity, up to small corrections arising from either the non-zero strange quark mass or from higher order QCD contributions. In contrast, certain New Physics (NP) scenarios may generate $b \to s\gamma$ operators of comparable size to the SM terms, but with exotic parity structure, significantly modifying the expected ratio of left- versus right-handed photons -- the photon polarization ratio. Measurement of this ratio therefore has the potential to test the parity structure of $b \to s\gamma$ operators against SM expectations, as well as either constrain or detect the signatures of such NP scenarios.

The $b \to s\gamma$ photon polarization ratio may be measured via various different approaches. Dominantly right-handed photon production in resonant $B \to (K_{1}(1400)  \to K \pi \pi)\gamma$ generates an up-down asymmetry of the photon momentum with respect to the $K_{1} \to K \pi\pi$ decay plane \cite{Gronau:2001ng,Gronau:2002rz}. This up-down asymmetry was recently measured by the LHCb collaboration~\cite{Aaij:2014wgo}. However, a theoretical prediction for the asymmetry is not yet available, so that the photon polarization ratio cannot yet be extracted from these results or compared to SM expectations. Along similar lines, the photon polarization ratio may also be probed by measuring the spin fraction of $\Lambda$'s in unpolarized $\Lambda_b \to \Lambda \gamma$ decays \cite{Mannel:1997pc}, or by measuring an angular asymmetry between the $\Lambda_b$ spin and the outgoing photon momentum for polarized $\Lambda_b \to \Lambda \gamma$ \cite{Hiller:2001zj}.  Other methods look for time dependent CP violation induced by mixing of $B \to K^* \gamma$ and $\bar{B} \to K^*\gamma$, which is proportional to the photon polarization ratio \cite{Atwood:1997zr,Kruger:1999xa}. Additionally one may probe the polarization ratio by looking for asymmetries in angular observables in resonant $B \to K \pi \ell^+\ell^-$ or $B \to \pi\pi \ell^+\ell^-$ \cite{Kruger:1999xa,Kim:2000dq}, or look for transverse asymmetries in the dilepton invariant mass for non-resonant $B \to K\pi \ell^+\ell^-$ \cite{Melikhov:1998cd}.

In this work we focus on the $B \to (K^* \to K^+\pi^-)\gamma$ process, in which the on-shell photon subsequently undergoes Bethe-Heitler (BH) nuclear conversion inside the detector to a lepton-antilepton pair. The cross-section for BH pair conversion of $\sim$ GeV photons is approximately two (eight) orders of magnitude larger than the Compton (Rayleigh) scattering cross-section (\cite{Agashe:2014kda}; see chapter 32), so that to an excellent approximation the emitted photon does not decohere before conversion. The $\bkg$ photon polarization ratio, $r$, is precisely defined via the amplitude ratio
\begin{equation}
	\label{eqn:DRPD}
	r e^{i(\phi + \delta)} \equiv \frac{\mathcal{A}(\bkg_L)}{\mathcal{A}(\bkg_R)}\,,
\end{equation}
in which $\phi$ ($\delta$) is a relative weak (strong) phase. In the SM, $r$ is expected to be at most $\sim\Lambda_{\rm qcd}/m_b$ \cite{Grinstein:2004uu}, while the weak phase $\phi$ is suppressed by $\Lambda_{\rm qcd}/m_b~|V_{ub}^{\phantom{*}}V_{us}^*/V_{tb}^{\phantom{*}}V_{ts}^*| \ll 1$. Hence measurement of not only $r$, as discussed above, but also $\phi$ may test SM expectations: Measuring either $r$ or $\phi \sim \mathcal{O}(1)$ would be highly suggestive of NP effects.

Measurement of $r$ via BH conversion was first considered in Ref.~\cite{Grossman:2000rk}. In that analysis, the $B$ meson was assumed to be at rest relative to the conversion nucleus. Further, the conversion itself was assumed be to a perfect linear polarizer of the photon, so that the conversion leptons and photon are constrained to be coplanar. However, in practice the $B$ meson is typically at least semi-relativistic, and typically an $\mathcal{O}(1)$ fraction of conversion events have non-negligible acoplanarity (see e.g. Ref.~\cite{Bishara:2013vya}). This leads to a richer phase space structure for the outgoing conversion leptons, kaon and pion. Moreover, interference effects between the $\bkg$ decay and the BH conversion amplitudes were not included. The key motivation to reconsider the above analysis, then, is to include these $B$-boost, acoplanarity and interference effects. We exploit recent compact results for BH conversion spin-helicity amplitudes \cite{Bishara:2013vya} to construct the full $\bfull$ amplitude, and show that interference between its decay and conversion components permits both $r$ and $\phi$ to be probed by the kinematic configuration of the final state conversion leptons, kaon and pion. In particular, we develop an angular observable that probes both of these parameters. This new analysis admits arbitrary boosts of the parent $B$ meson relative to the BH conversion nucleus, and includes lepton-photon acoplanarity, which turns out to play an important role in enhancing the $r$- and $\phi$-sensitive interference effects.

Performing an experiment to measure $r$ and $\phi$ with this technique will be challenging. In the first instance, precise reconstruction of the leptonic momenta is required, which can only be achieved, even in principle, if the leptonic opening angles are larger than the angular resolution of the detector. Typically, the leptonic opening angle after BH conversion is $\theta_{\ell\ell} \sim \me/E_\gamma \sim 10^{-4}$, for typical photon energies in a $\bkg$ decay with a semirelativistic $B$. Specifically, for a photon of energy $\lesssim 5$~GeV, the probability for $\theta_{\ell\ell} > 10^{-4}$ ($10^{-3}$) is $\sim 98$\% ($43$\%). Hence exquisite angular resolutions will be required. A further complicating factor is the multiple rescattering of the leptons in the detector material after conversion. The rms rescattering angle in matter is $\simeq (13.6~\mbox{MeV}/E_\ell) \sqrt{x/x_0}$ \cite{Agashe:2014kda}, where $x/x_0$ is the path length inside the detector in units of radiation length. For $x/x_0 \sim$ few \% -- a typical value -- the rescattering angle is comparable to the typical opening angle, $\theta_{\ell\ell}$. Finally, the probability of photon conversion itself is typically low at current and planned $B$ factories, being at most of order a few percent. This probability depends mildly on the detector design. For example, it is approximately 3\%, 2-3\%, and 6\% at BaBar~\cite{Boutigny:1995ib}, LHCb~\cite{Aaij:2014jba}, and Belle II~\cite{BelleII:2010td} respectively. For all these reasons, this technique will likely only be feasible with a dedicated detector element that has a large scattering length, e.g. a gaseous TPC \cite{Harpo:2013ha}. 

For these reasons, in this work we shall restrict ourselves to a thought experiment-type approach. That is, we develop explicit analytic expressions for amplitudes and observables with respect to the underlying $\bfull$ process alone, but do not include smearing from leptonic rescattering and limited kinematic resolution, or realistic detector simulations. Based on the results of this work, future studies may perhaps incorporate these latter effects.

\section{Amplitudes}
\subsection{Amplitude factorization}
Keeping operators up to dimension five, the effective theory of interest for $B \to K^* \gamma$ and $\bar{B} \to \bar{K}^*\gamma$ decays may be written in the general form
\begin{equation}
	\mathcal{L}_{\rm eff} = \gpl B^\dagger K^*_{\mu\nu} F^{\mu\nu} + \gpp B^\dagger K^*_{\mu\nu} \tilde{F}^{\mu\nu}  + \gbpl \bar{B}^\dagger \bar{K}^*_{\mu\nu} F^{\mu\nu} + \gbpp \bar{B}^\dagger \bar{K}^*_{\mu\nu} \tilde{F}^{\mu\nu}\,,
\label{eqn:LEFF}
\end{equation}
where $\tilde{X}^{\mu\nu} \equiv \epsilon^{\mu\nu\rho\sigma}X_{\rho\sigma}/2$, and the dimensionful couplings generically contain relative strong phases. We consider only $K^*$ decays to charged pseudoscalars, i.e. $K^* \to K^+\pi^-$ or $\bar{K}^* \to K^- \pi^+$. The sign of the pion or kaon charge therefore tags the $K^*$ versus the $\bar{K}^*$, and hence tags the parent meson as either a $B$ or $\bar{B}$, up to electroweak loop suppressed corrections.  Hence we neglect interference effects from $B$-$\bar{B}$ mixing. 

We assume the conversion nucleus is spin-$0$, e.g. a $^{28}$Si nucleus, which is the dominant silicon isotope. The external quantum numbers for the full $B \to (K^* \to K^+\pi^-) (\gamma \stackrel{\rm{BH}}{\to} e^+e^-)$ helicity amplitudes are then just the spins of the electron and positron, denoted $r$ and $s= 1,2$ respectively. It follows that the full helicity amplitudes
\begin{align}
	\mathcal{M}_{rs} 
	& = \qquad \parbox[c]{6cm} {\includegraphics[width = 6cm]{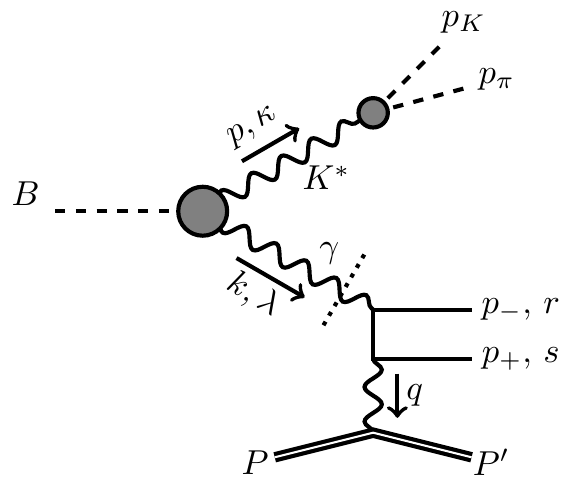}} + \quad \mbox{lepton exchanges}\notag\\
	& = i\frac{ g_{\mu\nu} - p_\mu p_\nu/\mk^2 }{p^2 - \mk^2 + i \mk \Gk} \sum_{\lambda = \pm}[\mcA_{\bkg}]^{\lambda\mu}[\mcA_{\kkp}]^{\nu}[\mcA_{\rm{BH}}]^{\lambda}_{rs}\,.
\end{align}
nearby to the $K^*$ Breit-Wigner peak. Here $\lambda = \pm$ ($\kappa =\pm,0$) is the helicity of the photon ($K^*$), and $k^\mu$ ($p^\mu$) is the photon ($K^*$) momentum; $P$ ($P'$) denotes the incoming (outgoing) nuclear momentum, with nuclear mass $P^2 = P^{'2} =  M^2$; $p_\pm$ ($p_{K,\pi}$) denote the momenta of the leptons (kaon and pion); and finally $q$ denotes the momentum exchange with the nucleus.

Momentum and angular momentum conservation in the $\bkg$ process ensure that $A_{\bkg}$ must annihilate the longitudinal component of the $K^*$ propagator. Applying the polarization completeness relation for $p^2 \not =0$, 
\begin{equation}
	\sum_{\kappa = \pm,0} \epsilon^\kappa_{\mu}(p) {\epsilon^{\kappa}}^*_{\nu}(p) = -g_{\mu\nu} + p_\mu p_\nu/ p^2\,, 
\end{equation}
we may factorize the full helicity amplitude into three helicity amplitude factors, 
\begin{subequations} 
\label{eqn:FAF}
\begin{equation}
	\mathcal{M}_{rs} = \frac{i}{p^2 - \mk^2 + i \mk \Gk} \sum_{\lambda, \kappa}[\mcA_{\bkg}]^{\lambda\kappa}[\mcA_{\kkp}]^{\kappa}[\mcA_{\rm{BH}}]^{\lambda}_{rs}\,.
\end{equation}
Similarly for the $\bar{B} \to (\bar{K}^* \to K^-\pi^+) (\gamma \stackrel{\rm{BH}}{\to} e^+e^-)$ process
\begin{equation}
	\overline{\mathcal{M}}_{rs} = \frac{i}{p^2 - \mk^2 + i \mk \Gk} \sum_{\lambda, \kappa}[\mcA_{\bbkg}]^{\lambda\kappa}[\mcA_{\kbkp}]^{\kappa}[\mcA_{\rm{BH}}]^{\lambda}_{rs}\,.
\end{equation}
\end{subequations}
Here and hereafter we neglect the mass splittings of the $B$-$\bar{B}$ and $K^*$-$\bar{K}^*$ systems, and denote the masses (momenta) of both CP conjugate states by $m_B$ and $\mk$ ($p_B$ and $p$) respectively.

\subsection{Kinematics}

The amplitude factors in eqs.~\eqref{eqn:FAF} are Lorentz invariants, and are naturally expressed with respect to kinematic coordinates that are defined in different frames. That is, the full $2 \to 5$ phase space -- the full coherent process is $B N \to K^+\pi^- \ell^+\ell^-N$, for nucleus $N$ --  is factored into two $1\to 2$ phase spaces, corresponding to $B \to K^*\gamma$ and $K^* \to K^+\pi^-$, and one $2 \to 3$ phase space for the BH conversion.  In general, a $2 \to 5$ phase space is fully specified by eleven coordinates. Two of these are the photon and $K^*$ invariant masses. The former is fixed to $k^2 = 0$ for an on-shell internal photon. With regard to the $K^*$ invariant mass, it is convenient to define hereafter the dimensionless quantity
\begin{equation}
	\sh \equiv p^2/\mk^2\,.
\end{equation}
In the narrow width limit, the Breit-Wigner factor
\begin{equation}
	\label{eqn:ZWL}
	\frac{1 }{|(\sh-1)\mk^2 + i \mk \Gamma_{K^*}|^2} \to \pi \frac{1}{\mk\Gamma_{K^*}} \delta[ (\sh-1) \mk^2]\,.
\end{equation}
That is, the narrow width limit corresponds to an on-shell $K^*$. Since, however, the $K^*$ has a finite width -- $\Gk/\mk \sim 5\%$ --  and need not be precisely on-shell, we shall treat $\sh$ as a phase space variable: The Breit-Wigner ensures $\sh$ is typically nearby to the $K^*$ mass shell up to the $K^*$ width, i.e. $\sh \simeq 1$ up to variations $\sim \Gk/\mk$.

\begin{figure}[h]
\parbox[c]{5cm}{\includegraphics[scale =1.25]{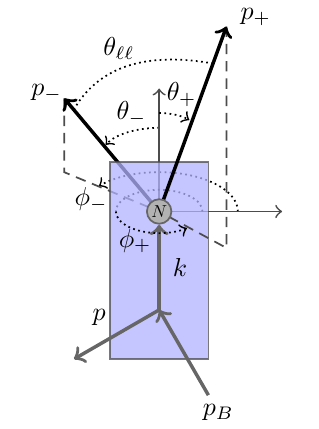}}  \parbox[c]{5cm}{\includegraphics[scale = 1.25]{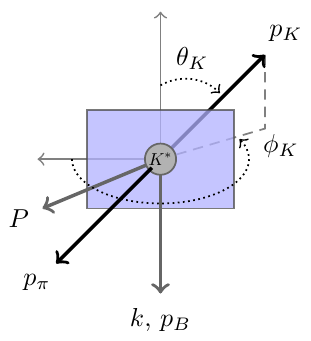}}  \parbox[c]{5cm}{\includegraphics[scale = 1.25]{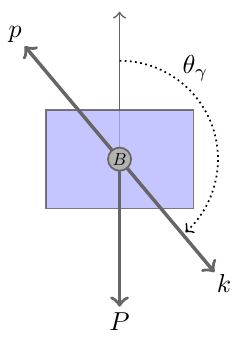}}
\caption{Kinematic configuration and coordinate choices.  $B$ momentum is denoted by $p_B$, and azimuthal angles are defined with respect to the $K^*$-$\gamma$ decay plane (blue). This plane contains $\bm{p}$ and $\bm{k}$ ($\bm{P}$ and $\bm{k}$) in the lab frame ($K^*$ or $B$ rest frames);  momenta lying in this plane in each frame are shown in gray. Left: Lepton polar angles $\theta_\pm$ and azimuthal angles $\phi_\pm$ in the lab frame. Middle: $\theta_K$ and $\phi_K$ polar angles in the $K^*$ rest frame. Right: The photon polar angle, $\theta_\gamma$, in the $B$ rest frame.}
\label{fig:DKC}
\end{figure}

In the case that the photon conversion material is cold, the lab frame coincides with the frame in which the BH conversion nucleus is at rest.  The following choices, shown in Fig.~\ref{fig:DKC}, for the remaining nine coordinates then prove convenient for the construction of compact and intuitive results: the lepton polar and azimuthal angles $\theta_\pm$ and $\phi_\pm$ and the energies $E_\pm$, defined in the nuclear rest frame -- the lab frame -- with respect to the photon momentum and the $K^*$-$\gamma$ decay plane, defined by $\bm{p}$ and $\bm{k}$;  the photon polar angle, $\theta_\gamma$, defined with respect to the nuclear momentum, $\bm{P}$, in the $B$ rest frame; the $K$ polar and azimuthal angles $\theta_K$ and $\phi_K$, defined in the $K^*$ rest frame with respect to the photon momentum, $\bm{k}$, and the plane defined by $\bm{P}$ and $\bm{k}$ in that frame. Note that the $K^*$-$\gamma$ decay plane is invariant under boosts between the lab, $K^*$ and $B$ rest frames, and therefore equivalent to the plane defined by $\bm{P}$ and $\bm{k}$ in either the $K^*$ or $B$ rest frames, as shown in Fig.~\ref{fig:DKC}.

\subsection{Helicity amplitude factors}
With these choices, we now proceed to explicitly compute the amplitude factors $[M_{\bkg}]^{\lambda\kappa}$, $[M_{\kkp}]^{\kappa}$, and $[M_{\rm{BH}}]^{\lambda}_{rs}$. Applying a light-cone decomposition to $p$, we define its associated null momentum with respect to the photon, i.e.
\begin{equation}
	\tilde{p}^\mu \equiv p^\mu - \frac{p^2 k^\mu}{2 p \cdot k}\,,
\end{equation}
and make the polarization gauge choices 
\begin{equation}
	\label{eqn:PGC}
	{\epsilon^\pm_{K^*}}^\mu(p) = \pm \frac{\sigbr{\mp}{k}{\sigma^\mu}{\tilde{p}}}{\sqrt{2}\asbr{k}{\mp}{\tilde{p}}{\pm}}\,, \qquad {\epsilon^\pm_{\gamma}}^\mu(k) = \pm \frac{\sigbr{\mp}{\bar{k}}{\sigma^\mu}{k}}{\sqrt{2}\asbr{\bar{k}}{\mp}{k}{\pm}} \,,
\end{equation}
for $\bar{k}$ an arbitrary null reference momentum. From the effective theory \eqref{eqn:LEFF} one may then show that the $\bkg$ and $\bbkg$ helicity amplitudes are
\begin{equation}
\begin{split}
	[\mcA_{\bkg}]^{\pm\pm} & = (\gpl \pm i \gpp)(\mb^2 - \mk^2\sh)\,,\\
	[\mcA_{\bbkg}]^{\pm\pm} & = (\gbpl \pm i \gbpp)(\mb^2 - \mk^2\sh)\,,
\end{split}
\end{equation}
and $[M_{\bkg}]^{\mp\pm} = [M_{\bkg}]^{\pm0} = 0$. Note that, by definition, $\mathcal{A}(\bkg_{R,L}) \equiv [M_{\bkg}]^{\pm\pm}$. It follows from eq. \eqref{eqn:DRPD} and its CP conjugate that
\begin{equation}
	\frac{\gpl - i\gpp}{\gpl + i \gpp} = re^{i(\delta + \phi)}\qquad \mbox{and} \qquad  \frac{\gbpl + i\gbpp}{\gbpl - i \gbpp} = re^{i(\delta - \phi)}\,.
\end{equation}
Hence
\begin{align}
	[\mcA_{\bkg}]^{++} & = (\gpl + i \gpp)(\mb^2 - \mk^2\sh)\,, \notag\\ 
	 [\mcA_{\bkg}]^{--}  & = re^{i(\delta + \phi)}(\gpl + i \gpp) (\mb^2 - \mk^2\sh)\,,\notag\\
	[\mcA_{\bbkg}]^{++} & = re^{i(\delta - \phi)}(\gbpl - i \gbpp) (\mb^2 - \mk^2\sh)\,, \notag\\
	 [\mcA_{\bbkg}]^{--}  & = (\gbpl - i \gbpp)(\mb^2 - \mk^2\sh)\,. \label{eqn:BKGHA}
\end{align}

The reference gauge momentum $\bar{k}$ in eqs. \eqref{eqn:PGC} is so far arbitrary. However, a particularly convenient choice is
\begin{equation}
	\label{eqn:KB}
	\bar{k}^\mu \equiv \frac{2 P\cdot k}{M^2} P^\mu - k^\mu\,,
\end{equation}
where $P^2 = M^2$ is the nuclear mass. In the lab frame -- the nuclear rest frame -- this choice \eqref{eqn:KB} ensures that for $k^\mu = E_\gamma(1,\hat{\bm{k}})$ then simply $\bar{k}^\mu = E_\gamma(1,-\hat{\bm{k}})$. We assume the nuclear scattering is coherent and quasi-elastic, i.e. that $P^{'2} = M^2$  -- equivalently $q^0 \equiv P{'^0} - P^0 = q^2/(2M)$ in the lab frame -- and that the outgoing nucleus is non-relativistic, so that the momentum exchange with the nucleus $|\bm{q}| \ll M$ (see Refs~\cite{RevModPhys.41.581,Tsai:1974py,Tsai:1977py} for a review of BH conversion). With these assumptions and the choice of $\bar{k}$ in eq.~\eqref{eqn:KB}, the BH spin-helicity amplitudes collapse to a simple form in the limit that the polar angles $\theta_\pm \ll 1$ and Lorentz factors $\gamma_\pm \equiv E_\pm/m \gg 1$, where $m$ is the lepton mass \cite{Bishara:2013vya}. At leading order in these limits,
\begin{equation}
	[\mcA_{\textrm{BH}}]^\lambda_{rs} \simeq e^3 \alpha^\lambda_{rs}\,,
\end{equation}
with
\begin{equation}
\begin{split}
	\alpha^-_{11}  = -(\alpha^+_{22})^* & =2 \sqrt{2\gamma_+\gamma_-} \frac{\sqrt{\mathcal{G}(q^2)}}{q^2} \bigg( \frac{1}{1 + \gamma_+^2\theta_+^2} - \frac{1}{1 + \gamma_-^2\theta_-^2}\bigg),\\
	\alpha^-_{^{12}_{21}}  =  +(\alpha^+_{^{21}_{12}})^* &  =  \pm  2 \sqrt{2\gamma_+\gamma_-} \frac{\sqrt{\mathcal{G}(q^2)}}{q^2} \frac{\gamma_{\mp}}{\gamma_+ + \gamma_-} \bigg( \frac{\gamma_+\theta_+ e^{-i\phi_+} }{1 + \gamma_+^2\theta_+^2} +  \frac{\gamma_-\theta_- e^{-i\phi_-} }{1 + \gamma_-^2\theta_-^2}\bigg),\\
	\alpha^-_{22}  = -(\alpha^+_{11})^* &  = 0 \,,\label{eqn:BHLOA}
\end{split}
\end{equation}
in which
\begin{equation}
	\label{eqn:QSLO}
	-q^2 \simeq m^2\Big(\gamma_+^2\theta_+^2 + \gamma_-^2\theta_-^2 + 2\gamma_-\gamma_+\theta_-\theta_+\cos(\phi_- - \phi_+)\Big) + \frac{m^2}{4}\bigg[\frac{1}{\gamma_+} + \frac{1}{\gamma_-}\bigg]^2\,.
\end{equation}
Here $\mathcal{G}(q^2)$ is the BH quasi-elastic form factor for the photo-nuclear vertex \cite{Tsai:1974py,Tsai:1977py}, 
\begin{equation}
	\mathcal{G}(q^2) = M^2 a^4q^4/(1 - a^2 q^2)^2\,,
\end{equation}
in which $a = 184.15(2.718)^{-1/2}Z^{-1/3}/m$ and $Z$ is the atomic number of the nucleus. This form factor encodes electronic screening of the nucleus, and regulates the $1/q^2$ pole in the amplitudes. From eqs.~\eqref{eqn:BHLOA}, one sees that the BH amplitudes are maximal at $\gamma_\pm\theta_\pm \sim 1$. That is, the typical lepton polar angle $\theta_\pm \sim m/E_\pm$. It also follows that the typical momentum exchange $-q^2 \sim m^2$, i.e. $|\bm{q}|^2 \ll M^2$ in concordance with our assumption of non-relativistic scattering.   

It now remains to compute the $\kkp$ and $\kbkp$ helicity amplitudes. Only transverse $K^*$ modes are generated by the $\bkg$ amplitude. Since $\epsilon_{K^*}^\pm \cdot p = 0$ and $p = p_K + p_\pi$, these amplitudes must therefore take the form
\begin{equation}	
	\label{eqn:AKKP}
	[\mcA_{\kkp}]^\kappa  = g_{K^*}{\epsilon_{K^*}^\kappa}_\mu (p) \big( p_K^\mu - p_\pi^\mu\big)\,,
\end{equation}
where $g_{K^*}$ is a dimensionless coupling, containing a strong phase. There are no other weak phases as $K^*$ decays strongly to $K\pi$. Under the polarization conventions \eqref{eqn:PGC}, and computing in the $K^*$ rest frame defined by Fig.~\ref{fig:DKC}, the helicity amplitudes are just spherical harmonics
\begin{equation}
	\label{eqn:KKPA}
	[\mcA_{\kkp}]^\pm(\theta_K, \phi_K) =\frac{e^{ \pm i\phi_K}}{\sqrt{2}}g_{K\pi}p_{K\pi}\sin\theta_K\,,
\end{equation}
with the momentum
\begin{equation}
	p_{K\pi} \equiv \frac{1}{\sh^{1/2}\mk}\big[\mk^2\sh - (m_K + m_\pi)^2\big]^{1/2}\big[\mk^2\sh- (m_K - m_\pi)^2\big]^{1/2}\,.
\end{equation}
Under CP, note that the amplitude transforms as
\begin{equation}
	[\mathrm{CP}~\mcA_{\kkp}]^\pm(\theta_K, \phi_K) = [\mcA_{\kbkp}]^\mp(\theta_K, \phi_K) = [\mcA_{\kkp}]^\pm(\theta_K, -\phi_K)\,.
\end{equation}
That is, defining $\phi_K$ and $\theta_K$ with respect to $\bar{K}^*$ rest frame just as for the $K^*$ in Fig.~\ref{fig:DKC}, then
\begin{equation}
	\label{eqn:KBKPA}
	[\mcA_{\kbkp}]^\pm(\theta_K, \phi_K) = \frac{e^{\pm i\phi_K}}{\sqrt{2}}\bar{g}_{K\pi}p_{K\pi}\sin\theta_K\,.
\end{equation}

\subsection{Full Amplitude}

Applying all the results \eqref{eqn:BKGHA}, \eqref{eqn:BHLOA}, \eqref{eqn:KKPA} and \eqref{eqn:KBKPA} to eqs. \eqref{eqn:FAF}, and defining $|\alpha|^2 \equiv \sum_{\lambda,r,s}|\alpha^\lambda_{rs}|^2$, the unpolarized square amplitudes
\begin{equation}
	\begin{split}
	|\mathcal{M}|^2 & =  A(r) \sin^2\theta_K \Big\{ |\alpha|^2 + \frac{8r}{1+r^2}\mbox{Re}\Big[\alpha^-_{12}\alpha^-_{21}e^{i(\phi +\delta- 2\phi_K)}\Big]\Big\}\,,\\
	|\overline{\mathcal{M}}|^2 & = \bar{A}(r)\sin^2\theta_{K}\Big\{ |\alpha|^2 + \frac{8r}{1+r^2}\mbox{Re}\Big[\alpha^-_{12}\alpha^-_{21}e^{i( \phi -\delta - 2\phi_K)}\Big]\Big\}\,,
	\end{split}
	\label{eqn:SAR}
\end{equation}
in which
\begin{equation}
	\begin{split}
	A(r) & \equiv \frac{e^3}{4}(1+r^2)\frac{|\gpl + i\gpp|^2|g_{K\pi}|^2}{|\mk^2(\sh-1) + i \mk \Gamma_{K^*}|^2} p^2_{K\pi}(\mb^2 -\mk^2\sh)^2\,,\\
	\bar{A}(r) & \equiv \frac{e^3}{4} (1+r^2)\frac{|\gbpl - i\gbpp|^2|\bar{g}_{K\pi}|^2}{|\mk^2(\sh - 1) + i \mk \Gamma_{K^*}|^2}p^2_{K\pi}(\mb^2 - \mk^2\sh)^2\,.
	\end{split}
\end{equation}
Eqs.~\eqref{eqn:SAR} compactly express the unpolarized square amplitude for the full $\bfull$ process in terms of just the BH conversion helicity amplitudes \eqref{eqn:BHLOA} and trigonometric (exponential) functions of the kinematic observables $\theta_K$ ($\phi_K$). The dependence on the parameters $r$, $\phi$ and $\delta$ is explicit and elementary.

As a cross-check of these results, we provide an alternative and more traditional derivation of the square amplitude in Appendix \ref{app:PD}, via construction of linearly polarized photon BH amplitudes. The consequent result~\eqref{eq:poldecsqamp} and the square amplitude~\eqref{eqn:SAR} are in excellent numerical agreement in the $\gamma_\pm \gg 1$ and $\theta_\pm \ll 1$ regime, applicable to the $\bfull$ process.  Note that the result~\eqref{eq:poldecsqamp} does not incorporate these approximations, so that the compact and explicit eqs.~\eqref{eqn:SAR} are strictly an approximation to eq.~\eqref{eq:poldecsqamp}. 

\section{Observables}
\subsection{Differential rate}
Making use of the explicit $r$ and $\phi$ dependence in the square amplitude results \eqref{eqn:SAR}, we may now proceed to extract $r$ and $\phi$ sensitive observables. First, however, we construct the full differential rate. The factorization \eqref{eqn:FAF} ensures that the phase space with an on-shell internal photon may be partitioned into a $B \to (K^* \to K\pi)\gamma$ cascade decay and a $\gamma N \to \ell^+\ell^-N$ conversion. That is, the differential rate for the full $\bfull$ process can be written as
\begin{equation}
	d\mathcal{R} = \big|\mathcal{M}\big|^2 d \mathcal{P}_B d\mathcal{P}_{\rm{BH}}
\end{equation}
where $d \mathcal{P}_B$ ($d\mathcal{P}_{\rm{BH}}$) is phase space factor of the decay (conversion). Note that $d\mathcal{R}$ here has the dimensions of a cross-section times a partial width. 

Each phase space factor is Lorentz invariant, and are naturally computed in different frames, as shown in Fig.~\ref{fig:DKC}. Computing in the $B$ rest frame followed by the $K^*$ rest frame, the phase space factor for the cascade decay is
\begin{align}	
	d\mathcal{P}_B  
	& = \frac{1}{2\mb} \frac{1}{(2\pi)^5} \frac{d^3p_\pi}{2E_\pi} \frac{d^3p_K}{2E_K}\frac{d^3 k}{2E_\gamma}  \delta^4(p_B -k- p_K -p_\pi)\,,\notag \\
	& \to \frac{\mk p_{K\pi}}{\sh^{1/2}} \frac{\mb^2 - \mk^2\sh}{32(2\pi)^4\mb^3}~d\Omega_K d\cos\theta_\gamma d\sh\,,
\end{align}
performing all trivial integrals, including over the overall azimuthal orientation of the $K^*$-$\gamma$-$N$ plane. Similarly, for the BH conversion, computing in the lab frame,
\begin{align}
	d\mathcal{P}_{\rm{BH}}
	& = \frac{1}{2M 2E_\gamma}\frac{1}{(2\pi)^5}\frac{d^3p_+}{2E_+} \frac{d^3p_-}{2E_-}\frac{d^3 P'}{2E'} \delta^4(P  + k - P' -p_+ -p_-)\,,\notag\\
	& \to \bigg[\frac{\mathcal{E}^2 - \Delta^2}{64 (2\pi)^5 M^2 \mathcal{E}} \bigg]d\Omega_+d\Omega_- d\Delta\,.
\end{align}
Here the lepton momenta been approximated in the measure via $\sqrt{E_\pm^2 - m^2} \simeq E_\pm$. We have further defined
\begin{equation}
	\label{eqn:EDD}
	\mathcal{E} \equiv \frac{E_+ + E_-}{2}\,,\qquad \Delta \equiv \frac{E_+ - E_-}{2}\,,
\end{equation}
and enforced non-relativistic nuclear scattering, which implies $E' \simeq M$ or equivalently $E_\gamma \simeq E_+ + E_-$, up to $q^0 = q^2/2M \ll m$ corrections. Hence to an excellent approximation $\mathcal{E}$ is half the photon energy in the lab frame. Moreover, note that $m\le E_\pm \le E_\gamma$ implies that
\begin{equation}
	\Delta \in (m-\mathcal{E}, \mathcal{E}-m)\,.
\end{equation}

At an $e^+e^-$ $B$-factory, such as Belle or BaBar, the $e^+e^- \to \Upsilon \to B\bar{B}$ production factorizes from the subsequent $B$ decays, because the $B$ is a pseudoscalar. In this type of collider, the rapidity of parent $B$ meson in the lab frame has a known prior probability distribution, $f_B(\eta) d\eta$, determined by the collider configuration, and enters as an independent phase space factor in $d\mathcal{R}$. In Appendix \ref{app:BR} we include a derivation of the $B$ rapidity pdf \eqref{eqn:BRPDF}, for an $e^+e^-$ machine. We shall restrict ourselves hereafter to the case that $f_B(\eta)$ is known. In this case, note that the energy $\mathcal{E}$ is fully specified by $\eta$, $\sh$ and $\theta_\gamma$, viz.
\begin{equation} 
	\mathcal{E}(\eta,\theta_\gamma,\sh) \equiv \frac{\mb^2 - \mk^2\sh}{4\mb}\big( \cosh \eta + \cos\theta_\gamma \sinh\eta\big)\,,
\end{equation}
so that the lepton energies can be expressed in terms of $\eta$, $\sh$, $\theta_\gamma$ and $\Delta$, via eqs. \eqref{eqn:EDD}. In our discussion of the kinematics above, it was convenient to express the amplitudes in terms of the ten phase space coordinates  $\sh$, $\theta_\gamma$, $E_\pm$, $\Omega_K$ and $\Omega_\pm$. We see here, however, that for the differential rate, it is more natural to choose $\eta$ and $\Delta$ as phase space coordinates rather than $E_\pm$.  Combining the above results together, the full differential rate 
\begin{equation}
	\label{eqn:DSG}
	d\mathcal{R} = \big|\mathcal{M}\big|^2 f_B(\eta) \frac{\mk p_{K\pi}}{\sh^{1/2}} \bigg[ \frac{\mb^2 - \mk^2\sh }{4(4\pi)^9\mb^3}\bigg]\bigg[\frac{\mathcal{E}^2(\eta,\theta_\gamma,\sh) - \Delta^2}{M^2 \mathcal{E}(\eta,\theta_\gamma,\sh)} \bigg] d\Omega_K d\Omega_+d\Omega_- d\Delta d\cos\theta_\gamma d\sh d\eta\,.
\end{equation}

\subsection{Polarization and weak phase observables}
Let us now define a further change of azimuthal angular coordinates, modulo $2\pi$
\begin{equation}
\begin{split}
	\psi & \equiv \phi_+ + \phi_- + 2\phi_K\,,\\
	\bar{\psi} & \equiv \phi_+ + \phi_- - 2\phi_K\,,\\
	\varphi & \equiv \phi_+ - \phi_-\,.
\end{split}
\end{equation}
The angle $\varphi$ encodes the acoplanarity of the leptons with respect to the photon, with coplanarity corresponding to $\varphi = \pi$. Note that $\phi_K$ and $\phi_\pm$ are defined with opposite orientations around the photon momentum direction $\hat{\bm{k}}$ (see Fig.~\ref{fig:DKC}). For coplanar conversion leptons and a stationary $B$ in the lab frame, $\psi$ then corresponds to the relative twist between the positron-electron conversion plane and the $K$-$\pi$ decay plane. Similarly, $\bar{\psi}$ would then correspond to the averaged orientation of the positron-electron conversion plane and the $K$-$\pi$ decay plane with respect to the $K^*$-$\gamma$ decay plane.

From eq. \eqref{eqn:QSLO}, the momentum exchange has the form $q^2 \propto 1 + \zeta \cos\varphi$, with $\zeta <1$. It follows from eqs. \eqref{eqn:BHLOA}, \eqref{eqn:SAR} and \eqref{eqn:DSG} that the differential rate can be written in the form
\begin{multline}
	d\mathcal{R} = \Big(\sum_k a_k \cos^k(\varphi) \Big)\Big[A_{1} + A_{2} \cos (\psi + \varphi - \phi - \delta) + \\
		A_3 \cos (\psi - \varphi - \phi - \delta) + A_4 \cos (\psi - \phi - \delta)\Big]\,,
\end{multline}
where $a_k$ and $A_i$ are purely functions of the phase space orthogonal to $\psi$, $\bar{\psi}$ and $\varphi$. That is, $a_k$ and $A_i$ are functions of $\sh$, $\eta$, $\theta_{\gamma,K,\pm}$ and $\Delta$. Integrating over all phase space except $d\psi$, we see that the marginal differential rates for $\bfull$ and $\bbarfull$ must respectively have the form
\begin{equation}
\begin{split}
	\frac{d\mathcal{R}}{d\psi} & = \frac{\mathcal{R}}{2\pi}\Big[1 - \frac{2r}{1+r^2} \mcC\cos(\psi -\phi - \delta)\Big]\,,\\
	\frac{d\overline{\mathcal{R}}}{d\psi} & = \frac{\overline{\mathcal{R}}}{2\pi}\Big[1 - \frac{2r}{1+r^2} \mcC\cos(\psi -\phi + \delta)\Big]\,. \label{eqn:MDR}
\end{split}
\end{equation}
Eqs. \eqref{eqn:SAR} tell us that the cosine coefficient $\mcC$ arises from a ratio of BH interference terms to the BH squared amplitude, and is therefore independent of $r$ or $\phi$. In other words, this coefficient is the same for both of the CP conjugate processes, and is $\bkg$ operator independent. We have chosen the relative sign of $\mcC$ in eqs~\eqref{eqn:MDR} to anticipate the choice that ensures $\mcC >0$. Further, we have chosen the normalization of $\mcC$ in eqs~\eqref{eqn:MDR} to ensure, via positive semi-definiteness of $d\mathcal{R}/d\psi$, that $|\mcC| \le 1$, noting that the $r$ dependent factor $2r/(1+r^2) \le 1$ for any $r$. The coefficient $\mcC$ may then be interpreted as the maximum possible ratio of the amplitude of $d\mathcal{R}/ d\psi$ oscillations to their average value, $\mathcal{R}/2\pi$. Hereafter we call this ratio the relative oscillation amplitude. 

Eqs~\eqref{eqn:MDR} are the main results of this paper. Once the coefficient $\mcC$ is computed, then measurement of the relative oscillation amplitude in $d\mathcal{R}/d\psi$ permits extraction of $r$ up to the two-fold ambiguity $r \leftrightarrow 1/r$. Further, measurement of the average phase shift (phase shift difference) between $d\mathcal{R}/d\psi$ and $d\overline{\mathcal{R}}/d\psi$ permits extraction of the weak (strong) phase $\phi$ ($\delta$). Equivalently, one may construct two forward-backward type asymmetries. Defining the four quadrants $\mathrm{I}: \psi \in [0,\pi/2]$, $\mathrm{II}: \psi \in [\pi/2,\pi]$, $\mathrm{III}: \psi \in [\pi,3\pi/2]$ and $\mathrm{IV}: \psi \in [3\pi/2,2\pi]$ then
\begin{subequations}
\label{eqn:SIGOM}
\begin{equation}
\begin{split}
	\Psi_{\psi} & \equiv \mathcal{R}^{-1} \int_{-\mathrm{I} - \mathrm{II} + \mathrm{III} + \mathrm{IV}} \frac{d\mathcal{R}}{d\psi} d\psi = \frac{2}{\pi}\frac{2r}{1+r^2}\mcC\sin (\phi +\delta)\\
	\Omega_{\psi} & \equiv \mathcal{R}^{-1}  \int_{-\mathrm{I} + \mathrm{II} + \mathrm{III} - \mathrm{IV}} \frac{d\mathcal{R}}{d\psi} d\psi = \frac{2}{\pi}\frac{2r}{1+r^2}\mcC\cos (\phi +\delta)\,,
\end{split}
\end{equation}
and moreover
\begin{equation}
\begin{split}
	\overline{\Psi}_{\psi} & \equiv \overline{\mathcal{R}}^{-1} \int_{-\mathrm{I} - \mathrm{II} + \mathrm{III} + \mathrm{IV}} \frac{d\overline{\mathcal{R}}}{d\psi} d\psi =  \frac{2}{\pi}\frac{2r}{1+r^2}\mcC\sin(\phi - \delta)\\
	\overline{\Omega}_{\psi} & \equiv \overline{\mathcal{R}}^{-1} \int_{-\mathrm{I} + \mathrm{II} + \mathrm{III} - \mathrm{IV}} \frac{d\overline{\mathcal{R}}}{d\psi} d\psi =  \frac{2}{\pi}\frac{2r}{1+r^2}\mcC\cos(\phi - \delta)\,.
\end{split}
\end{equation}
\end{subequations}
Note that all four symmetries have an upper bound $2/\pi$. For known $\mcC$, one may extract $r$ and $\phi \pm \delta$ from these two sets of asymmetries.

\subsection{Statistics and Sensitivity Enhancements}
Before proceeding to numerical computation of $\mcC$, let us pause to consider the statistical confidence in the extraction of $r$ and $\phi$. We focus on their extraction from the asymmetries \eqref{eqn:SIGOM}. These asymmetries are expectation values of a random variable defined to take the values $\pm 1$ on the quadrants I, II, III and IV, as specified in eqs. \eqref{eqn:SIGOM}. For a sample of $N\gg1$ events, the corresponding error $\sigma_X = \sqrt{(1 - X^2)/N}\simeq 1/\sqrt{N}$, for $X = \Psi_{\psi}$, $\Omega_{\psi}$, $\overline{\Psi}_{\psi}$, and $\overline{\Omega}_{\psi}$. The statistical confidence at which one rejects the SM values $X_{\textrm{SM}}$ -- thus measuring NP effects -- is then characterized by the chi-square statistic $(X - X_{\rm SM})^2/\sigma_X^2$.

As shown in Ref.~\cite{Bishara:2013vya} and below, special `sensitivity parameter' kinematic cuts may enhance $\mcC$ on the resulting remaining phase space. The construction of these cuts is motivated by the observation, from eqs.~\eqref{eqn:SAR}, that $\mcC$ is enhanced on those areas of phase space in which the BH interference term, $\sim \alpha^-_{12}\alpha^-_{21}$, is comparable or larger than terms in the total BH square amplitude $|\alpha|^2 = \sum_{\lambda,r,s} |\alpha^\lambda_{rs}|^2$. For example, one may define the sensitivity parameter
\begin{align}
	\label{eqn:SCP}
	\mcS 
	&\equiv |\alpha^-_{12}\alpha^-_{21}|/|\alpha^-_{11}|^2\notag\\
	& \simeq 2(1-\cos[|\phi_+-\phi_-| - \pi])\bigg[\frac{\gamma_+\gamma_-}{(\gamma_+ + \gamma_-)^2}\bigg]
	\bigg[\frac{ \gamma_+ \theta_+ \gamma_-\theta_- }{(\gamma_+^2\theta_+^2 - \gamma_-^2\theta_-^2)^2} \bigg] (1 + \gamma_+^2\theta_+^2)(1+ \gamma_-^2\theta_-^2)\,.
\end{align}
Requiring $\mcS \gtrsim 1$ produces an event-level kinematic cut that typically leads to $\mathcal{O}(1)$ enhancements of $\mcC$ on the remaining phase space, as will be verified below. Note that the $(1-\cos[|\phi_+-\phi_-| - \pi])$ factor in eq.~\eqref{eqn:SCP} implies that $\mcS \simeq 0$ for coplanar events, and moroever that a lower bound on $\mcS$ typically favors events with higher acoplanarity. One may also consider other sensitivity parameters, such as
\begin{equation}
	\mcT \equiv 2 |\alpha^-_{12}\alpha^-_{21} |/ |\alpha|^2\,,
\end{equation}
which is normalized such that $\mcT\in[0,1]$.

Let us define $\mcC_0$ ($N_0$) to be the relative oscillation amplitude (number of events) in the absence of $\mcS$ or $\mcT$ cuts, and write
\begin{equation}
	\mcCc \equiv \mcC[\mcS > \mcSc, \mcT > \mcTc]\,,\qquad  \Nc \equiv N[\mcS > \mcSc, \mcT > \mcTc]\,.
\end{equation}
Compared to the $\mcSc = \mcTc = 0$ case, the application of sensitivity parameter cuts scales the NP statistical confidence by the factor 
\begin{equation}
	\label{eqn:DSS}
	\Sigma  \equiv \left(\frac{X-X_{\textrm{SM} }}{\sigma_X}\right)^2\bigg|_{\mcS > \mcSc, \mcT > \mcTc} \left(\frac{\sigma_{X}}{X - X_{\textrm{SM}}}\right)^2 \simeq \left(\frac{\mcCc}{\mcC_0}\right)^2\frac{\Nc}{N_0}\,,
\end{equation}
for $X = \Psi_{\psi}$, $\Omega_{\psi}$, $\overline{\Psi}_{\psi}$, and $\overline{\Omega}_{\psi}$. That is, the enhancement $\mcCc/\mcC_0$ achieved by the sensitivity parameter cuts competes with the corresponding increase in the statistical error, $\sqrt{N_0/\Nc}$, since necessarily $\Nc < N_0$. We shall see in the next section that there are choices of $\mcSc$ and $\mcTc$ for which $\Sigma > 1$. For the purpose of measuring NP effects, this is equivalent to an effective increase in the sample size $N_0 \mapsto \Sigma N_0$ -- an increase in the effective statistics. 

\section{Simulations}
\subsection{Relative Oscillation Amplitude}
Extraction of the relative oscillation amplitude coefficient $\mcC$ is achieved numerically via Monte-Carlo (MC) generation of $\bfull$ events according to the differential rate~\eqref{eqn:DSG}. We use the matrix element in eq.~\eqref{eq:poldecsqamp}, generated from linearly polarized photon amplitudes. Though it does not provide the same analytical insight as the matrix element generated from spinor-helicity methods~\eqref{eqn:SAR}, this matrix element is as numerically stable as the latter, and moreover, acts as a convenient cross-check of the analytic results in eqs.~\eqref{eqn:SAR} and \eqref{eqn:MDR}.  

For the numerical results shown in this paper, we use a private MC code written in \texttt{C/Python}. For simplicity, we apply the narrow $K^*$ limit~\eqref{eqn:ZWL}, which fixes the $K^*$ to be on-shell. We assume the nominal Belle II parameters \cite{BelleII:2010td} (see also App.~\ref{app:BR}) in order to determine the $B$ rapidity distribution $f_B(\eta)$, which is peaked at $\beta\gamma \simeq 0.29$. More details of the operation of this MC generator are included in Appendix~\ref{app:MC}. We have further checked the numerical results with a second private MC, written in \texttt{C++/Java}, that makes use of the matrix element~\eqref{eqn:SAR}. In both codes, we discard the overall normalization of the matrix element -- e.g. $A(r)$ in eqs.~\eqref{eqn:SAR} -- since we are concerned only with the relative oscillation amplitude, $\sim \mcC$.

\begin{figure}
\includegraphics[width=0.49\linewidth]{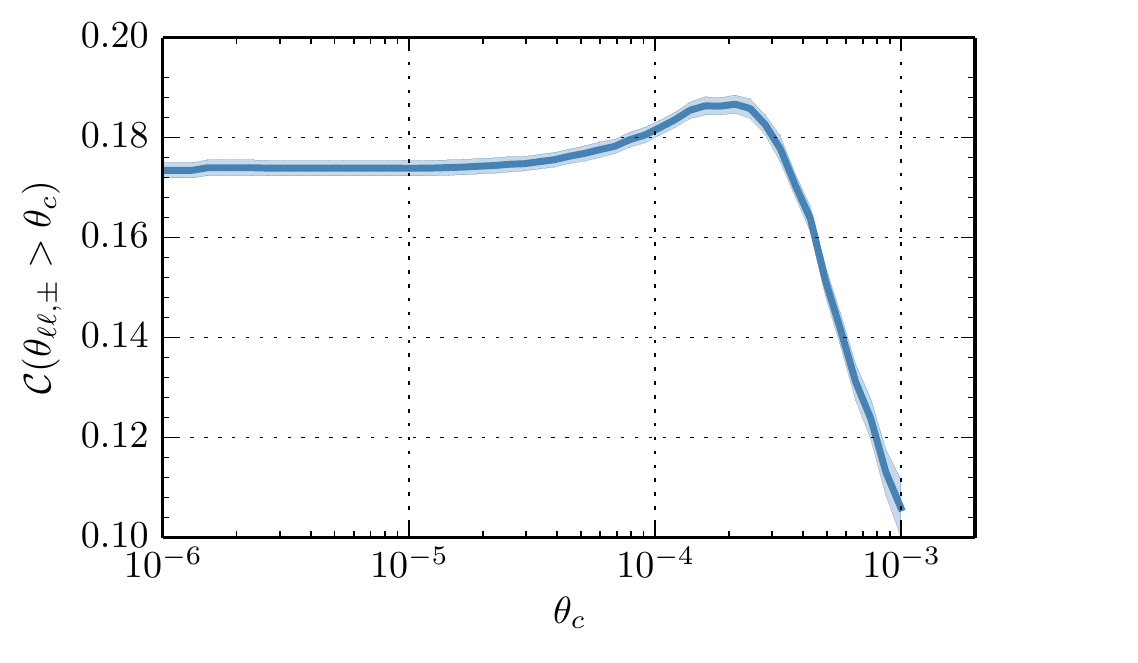}
\includegraphics[width=0.49\linewidth]{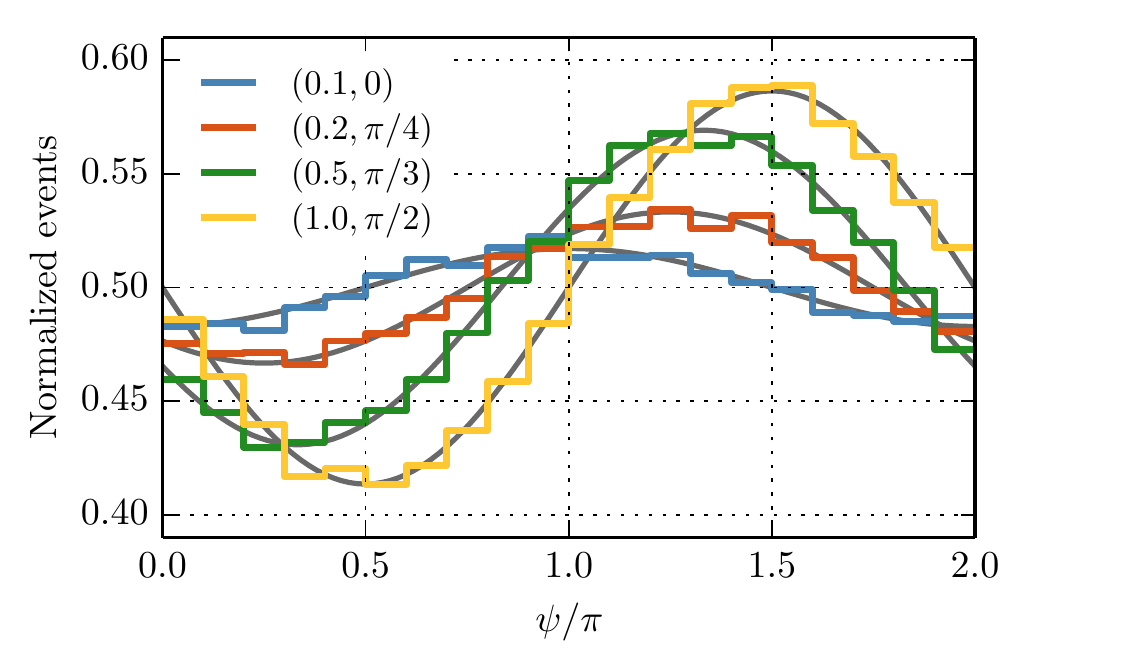}
\caption{Left: The fit value for $\mcC$ with the $\pm1\sigma$ error band as a function of the polar angle cuts $\theta_{\ell\ell,\pm} > \theta_{\textrm{c}}$ (see Fig.~\ref{fig:DKC}). The peak value of $\mcC$ approximately coincides with the peak of the $\theta_\pm$ marginal distribution (see the left panel of Fig.~\ref{fig:MC2}). Right: Normalized differential distribution $d\mathcal{R}/d\psi$ for four different $(r,\phi+\delta$) couplets and $\theta_{\textrm{c}} = 10^{-6}$. Also shown are theory predictions (gray) for the input values of $(r,\phi+\delta)$ and the extracted value $\mcC[\theta_{\textrm{c}} = 10^{-6}]$ in eq.~\eqref{eqn:EXC}.}
\label{fig:CVTH}
\end{figure}

In order to account for limited angular resolution, we include hereafter cuts on the lepton polar angles, $\theta_\pm$, and opening angle, $\theta_{\ell\ell}$, defined in Fig.~\ref{fig:DKC}. We will consider a uniform polar cut
\begin{equation}
	\theta_{\ell\ell, \pm} > \theta_{\textrm{c}}\,,
\end{equation}
for various values of $\theta_{\textrm{c}}$. In particular, we consider two benchmark cases $\theta_{\textrm{c}} = 10^{-6}$ and $5\times 10^{-4}$. The former captures almost all conversion leptons in the $\bfull$ process for semirelativistic $B$'s, while the latter might be plausibly achievable in the near- to mid-term future. To extract $\mcC$, we fit eq.~\eqref{eqn:MDR} to the $d\mathcal{R}/d\psi$ histograms for various choices of $r$ and $\phi+\delta$, including the couplets $\{r,\phi+\delta\}=\{(0.1,0),(0.2,\pi/4),(0.5,\pi/3),(1.0,\pi/2)\}$. In Fig.~\ref{fig:CVTH} we show the extracted $\mcC$ as a function of the $\theta_{\textrm{c}}$ cuts. The maximal relative oscillation amplitude one can expect is of $\mathcal{O}(20\%)$, and the benchmark extracted $\mcC$ values are
\begin{equation}
	\label{eqn:EXC}
	\mcC[\theta_{\textrm{c}} = 10^{-6}] = 0.173  \pm\, 0.001, \quad \mbox{and} \quad \mcC[\theta_{\textrm{c}} = 5\times10^{-4}] = 0.150\pm 0.003\,,
\end{equation}
where the errors are purely statistical in origin. We also show in Fig.~\ref{fig:CVTH} typical $d\mathcal{R}/d\psi$ histograms for various choices of $r$ and $\phi +\delta$. The expected shifted and amplitude-modulated cosine can be clearly seen. Applying the extracted value for $\mcC$, these histograms are in excellent agreement with the theory predictions \eqref{eqn:MDR}.

\subsection{Statistics Enhancements}
\begin{figure}[t]
\includegraphics[scale=0.9]{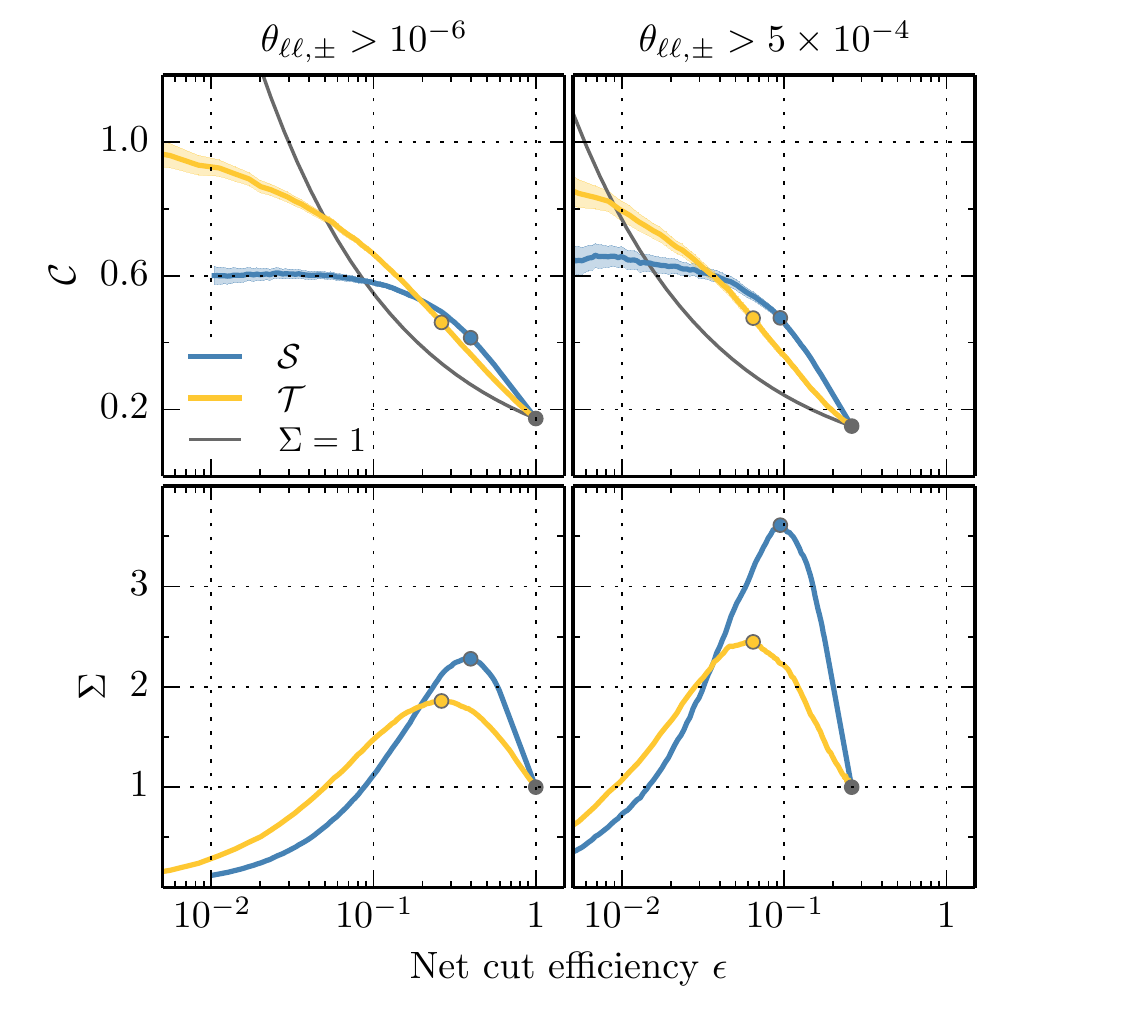}
	\caption{\emph{Upper panels:} The coefficient $\mcC$ as a function of the cut efficiency for the $\mcS$ (blue) and $\mcT$ (gold) kinematic cuts, with polar cuts $\theta_{\textrm{c}} = 10^{-6}$ (left) and $\theta_{\textrm{c}} = 5\times10^{-4}$ (right). The colored regions depict the $\pm 2\sigma$ statistical error bands.  The equivalent effective statistics curve $\Sigma = 1$, i.e. $\mcC = \mcC_0/\sqrt{N/N_0}$, is also shown (gray). \emph{Lower panels}: Statistics enhancement $\Sigma$ as a function of the cut efficiency. The maxima correspond to the optimum cuts $\mcS > \mcSc^{\textrm{opt}}$ and $\mcT >\mcTc^{\textrm{opt}}$ (colored dots in all panels).}
	\label{fig:STC}
\end{figure}

\begin{figure}[t]
\includegraphics[scale=0.9]{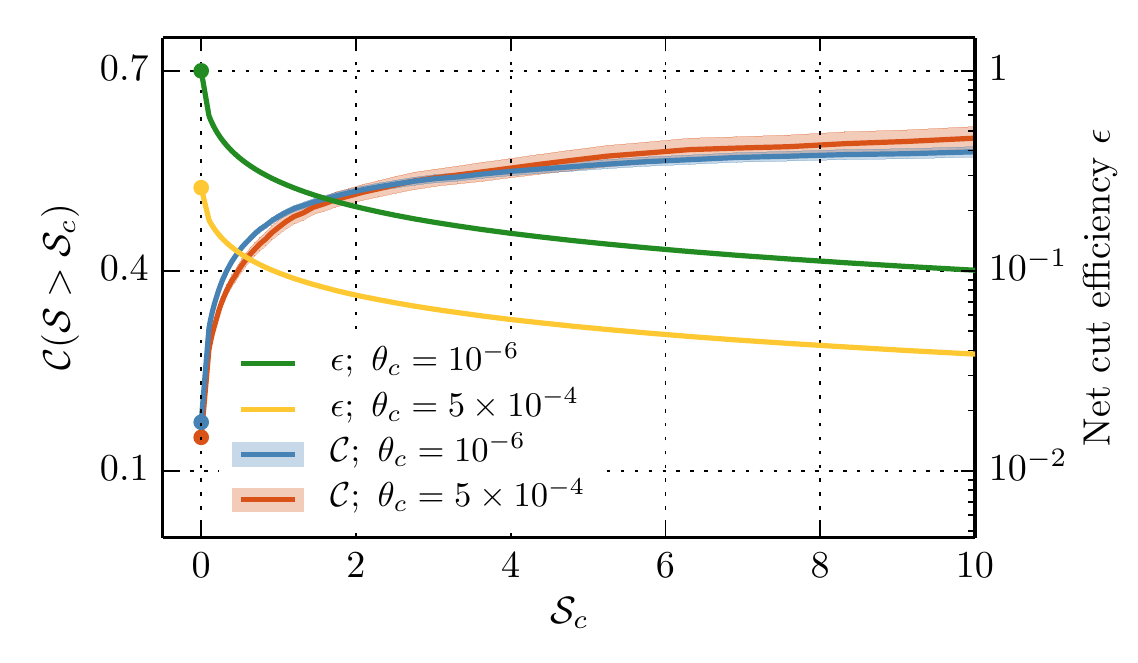}
	\caption{The enhancement in $\mcC$ as a function of $\mcS$. The secondary $y$-axis shows the corresponding cut efficiency. The colored regions depict the $\pm 2\sigma$ statistical error bands.}
	\label{fig:SC}
\end{figure}

Incorporating the $\mcS$ and $\mcT$ kinematic cuts, we show in Fig.~\ref{fig:STC} the absolute enhancement of $\mcC$ as a function of the net cut efficiency, $\epsilon$, for the two benchmark polar angle cuts. The net cut efficiency is defined hereafter to be the fraction of events kept after application of both kinematic and polar cuts. The pure $\mcS$ cuts -- that is, $\mcC(\mcS> \mcSc, \mcT >0)$ --  provide the larger enhancement at high cut efficiencies. At lower efficiencies the pure $\mcT$ cut provides the larger enhancement. The $\mcC$ and $\epsilon$ dependence on $\mcSc$ may be read off from Fig.~\ref{fig:SC}.

Comparing to the equivalent effective statistics curve $\Sigma = 1$ -- i.e. $\mcC = \mcC_0/\sqrt{N/N_0}$ --  we see in Fig.~\ref{fig:STC} that the $\mcC$ dependence initially rises faster than $\mcC_0/\sqrt{N/N_0}$. This means that for low values of $\mcSc$ and $\mcTc$, the statistics on the $\mcS$ and $\mcT$ cut phase space is enhanced compared to the full data set. The explicit $\Sigma$ dependence is also shown in Fig.~\ref{fig:STC}. In particular, one sees that for $\theta_{\textrm{c}} = 5\times10^{-4}$, the statistical enhancement $\Sigma$ is optimized at $\mcS^{\textrm{opt}}_{\textrm{c}} = 1.1$ ($\mcT^{\textrm{opt}}_{\textrm{c}}= 0.62$) corresponding to $\Sigma = 3.6$ ($\Sigma = 2.5$).

To demonstrate the efficacy of these statistics enhancements, in Table~\ref{tab:EXT} we compute the extracted values for $r$ and $\phi+\delta$, along with their statistical errors, with and without the optimal $\mcSc^{\textrm{opt}}$ and $\mcTc^{\textrm{opt}}$ cuts. We use an MC sample generated from an input couplet $(r,\phi + \delta) = (0.2,\pi/4)$, containing a total of $N=10^4$ events. This roughly corresponds to $50~\textrm{ab}^{-1}$ of data --  a benchmark luminosity at Belle II \cite{BelleII:2010td} -- and a percent level photon conversion rate with ideal acceptance efficiency.  The polar cut is $\theta_{\textrm{c}} = 5\times 10^{-4}$. The extracted values are obtained by two different methods, first from a fit to the differential rate eq.~\eqref{eqn:MDR} and second from the quadrant asymmetries \eqref{eqn:SIGOM}.

\begin{table}[h]
\def\arraystretch{1.3}
\newcolumntype{C}{ >{\centering\arraybackslash $} m{2.2cm} <{$}}
\resizebox{0.9\linewidth}{!}{
\begin{tabular*}{1.1\linewidth}{@{\extracolsep{\fill}}|c|CC|CC|CC|}
\hline
	 \multirow{2}{*}{Method} &  \multicolumn{2}{c|}{$\mcS,\mcT>0$} & \multicolumn{2}{c|}{$\mcS> \mcSc^{\textrm{opt}}$} & \multicolumn{2}{c|}{$\mcT > \mcTc^{\textrm{opt}}$}\\[-10pt]
	&  r & (\phi +\delta)/\pi  & r & (\phi +\delta)/\pi  & r & (\phi +\delta)/\pi  \\
\hline\hline
	$d\mathcal{R}/d\psi$ & 0.194 \pm 0.017 & 0.255 \pm 0.026 & 0.194 \pm 0.010 & 0.263 \pm 0.015 & 0.190 \pm 0.013 & 0.247 \pm 0.020 \\
	$\Psi_\psi$, $\Omega_\psi$ & 0.217 \pm 0.005 & 0.216 \pm 0.157 & 0.203 \pm 0.003 & 0.248 \pm 0.088 & 0.192 \pm 0.003 & 0.232 \pm 0.112 \\
	\hline
\end{tabular*}
}
\caption{Extracted values of $r$ and $(\phi+\delta)$ from an MC sample with input values $(0.2,\pi/4)$.}
\label{tab:EXT}
\end{table}

The statistical errors for the optimized kinematic $\mcS$ and $\mcT$ cuts are a $\mathcal{O}(1)$ factor smaller than for the full data set, as expected from the above numerical analysis: Application of these sensitivity parameter cuts improves the statistical power of the $r$ and $\phi$ extraction from $\bfull$ events. Moreover, even without these enhancements, both $r$ and $\phi$ are extracted with sufficient numerical precision to probe NP effects at the $r \sim 1$ or $\phi \sim 1$ level.

\section{Conclusions}
In this paper we have presented the helicity amplitudes and differential rate for the $\bkg \to K^+\pi^-\gamma$ process, in which the photon undergoes subsequent nuclear conversion to a lepton pair. Interference between the intermediate, on-shell photon polarizations in the coherent $\bfull$ process produces oscillations in the angular kinematic observable $\psi$. Measuring the amplitude and phase of these oscillations -- or equivalently two quadrant-type asymmetries $\Psi_\psi$ and $\Omega_\psi$ -- permits extraction in principle of the polarization ratio, $r$, and the relative weak phase, $\phi$, of the right- and left-handed $\bkg$ amplitudes. In this manner, SM expectations for both $r$ and $\phi$ may be tested.

We have employed private Monte Carlo simulations to compute the $\psi$ distribution and asymmetries as a function of $r$, $\phi$ and kinematic cuts. In particular, kinematic cuts with respect to the sensitivity parameters $\mcS$ and $\mcT$ may sufficiently amplify these oscillations, such that the overall statistical power of the $r$ and $\phi$ measurement is increased by an $\mathcal{O}(1)$ factor.

Implementing this approach using converted photons will be experimentally challenging, not least because of the high angular resolution required to reconstruct the conversion lepton kinematics. Moreover, a detector whose thickness is on the order of one radiation length or less is required to avoid multiple leptonic rescatterings, that otherwise smear the lepton kinematics. Nonetheless, the theoretically clean nature of the $r$- and $\phi$-sensitive observables presented in this work may perhaps encourage the use of this technique in a future dedicated detector element.

\acknowledgements
The authors thank Yuval Grossman, Roni Harnik, Jacques Lefrancois, Zoltan Ligeti, Marie-H\'el\`ene Shune, and Jure Zupan for helpful discussions. 
The work of FB is supported in part by the Fermilab Fellowship in Theoretical Physics and by the University of Cincinnati physics department Mary J. Hanna fellowship. Fermilab is operated by Fermi Research Alliance, LLC under Contract No. DE-AC02-07CH11359 with the United States Department of Energy. The work of DR is supported by the NSF under grant No. PHY-1002399.

\clearpage
\appendix
\section{The \texorpdfstring{$B\rightarrow K\pi e^+e^-$}{BKee} squared matrix element by polarization decomposition}
\label{app:PD}

The effective Lagrangian in eq. \eqref{eqn:LEFF} gives the following Feynman rule for the $BK^*\gamma$ vertex:
\begin{equation}
\parbox[c]{4.5cm}{\includegraphics[width=4.5cm]{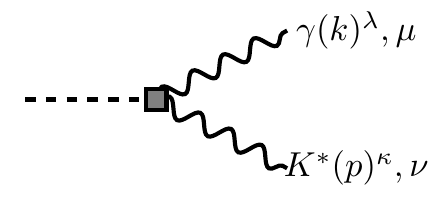}} = i\left[ \gpl\left( \Delta m_{BK^*}^2 g_{\mu\nu} - 2k_\nu p_\mu \right) + 2 \gpp \epsilon_{\mu\nu\rho\sigma}k^\rho p^\sigma \right]
\label{eq:feynrule}
\end{equation}
\noindent where $\Delta m_{BK^*}^2 \equiv m_B^2 - m_{K^*}^2$.
The amplitude for $B\rightarrow\gamma^\lambda K\pi$ with $\lambda$ being the photon helicity is then given by
\begin{equation}
\begin{split}
\mathcal{N}_\lambda = \gks\,\Big\{
  &\gpl\left[ \Delta m_{BK^*}^2\varepsilon_{\lambda}^*\cdot(p_K-p_\pi) - 2\,\varepsilon_{\lambda}^*\cdot p\,\left[ k\cdot(p_K-p_\pi) \right] \right] \\
	&\qquad-2\gpp \epsilon^{\mu\nu\rho\sigma}\left(p_K-p_\pi\right)_\mu\varepsilon^{\lambda*}_\nu k_\rho \,p_\sigma
\Big\}
\end{split}
\label{eq:bgkpiamp}
\end{equation}
The BH squared amplitude in the nuclear rest frame for a linearly polarized photon in the $+\hat z$ direction with polarizations $\lambda=\{1,2\}$ is
\begin{equation}
\begin{split}
\text{\rm{BH}}^{\lambda\lambda'} \simeq \frac{2e^6\mathcal{G}}{q^4}\left\{
		\frac{g^{\lambda\lambda'}\left[E_\gamma^2\,q^2+(k\cdot p_- + k\cdot p_+)^2\right]}{(k\cdot p_-)(k\cdot p_+)}
		-4\left(\frac{E_{p_+}p_-^\lambda}{k\cdot p_-}+\frac{E_{p_-}p_+^{\lambda}}{k\cdot p_+}\right)
			\left(\frac{E_{p_+}p_-^{\lambda'}}{k\cdot p_-}+\frac{E_{p_-}p_+^{\lambda'}}{k\cdot p_+}\right)
	\right\}\,,
\end{split}
\label{eq:bh}
\end{equation}
where terms of $\mathcal{O}(q^2/E_\pm^2)$ were dropped (see Appendix C in~\cite{Bishara:2013vya} for details). The squared amplitude is then given by
\begin{equation}
\left|\mcM\right|^2 = \sum_{\lambda,\lambda'\in\{1,2\}}\,\mcN_\lambda\mcN^*_{\lambda'}\text{\rm{BH}}^{\lambda\lambda'}.
\label{eq:poldecsqamp}
\end{equation}
A numerical comparison between the above expression and eq.~\eqref{eqn:SAR} shows excellent agreement over the entire phase space (sampled uniformly).

\section{\texorpdfstring{$B$}{B} rapidity distribution}
\label{app:BR}

Consider an $e^+e^- \to \Upsilon \to B\bar{B}$ factory, and let $\Theta$ denote the polar angle of the $B$'s with respect to the electron beamline in the center of mass frame. The amplitude for production $\mathcal{M}_{\rm prod} \sim \sin\Theta$, and so the probability distribution
\begin{equation}
	p_\Omega(\cos\Theta) = \frac{3}{4}\big[1 - \cos^2\Theta\big] \,.
\end{equation}
Here and in the following we neglect effects of lab frame angular acceptance cuts, which may non-trivially restrict the domain of both $\eta$ and $\Theta$. 

In the center of mass frame -- the rest frame of the $\Upsilon$ -- each $B$ has energy $E^* = m_\Upsilon/2$: hereafter the $^*$ superscript denotes center-of-mass frame quantities. The corresponding rapidity, which we choose to be positive by convention on the branch $\Theta \in [0,\pi]$,
\begin{equation}
	\eta^* = \cosh^{-1}(m_\Upsilon/ 2m_B)\,,
\end{equation}
and the $B$ speed in this frame $\beta^* = \tanh(\eta^*)$.

At $B$-factories the lab frame electron and position beam energies, $\mathcal{E}_\pm$, are asymmetric, but are chosen such that the $\Upsilon$ is on shell, i.e. $4\mathcal{E}_+\mathcal{E_-} = m_\Upsilon^2$. For example, at Belle II the beams are planned to be $\mathcal{E}_+ = 7$~GeV and $\mathcal{E}_- = 4$~GeV. The boost rapidity of the center of mass frame with respect to the lab frame is correspondingly
\begin{equation}
	\eta_\Upsilon = \cosh^{-1}[(\mathcal{E}_+ + \mathcal{E}_-)/m_\Upsilon]\,.
\end{equation}
The rapidity of the $B$ in the lab frame may now be written as a function of  $\cos\Theta$, viz.
\begin{equation}
	\eta(\cos\Theta) = \cosh^{-1}\Big\{ \cosh \eta^*\big[\cosh \eta_\Upsilon + \beta^*\cos\Theta \sinh \eta_\Upsilon\big]\Big\}\,,
\end{equation}
and its pdf, by definition
\begin{equation}
	f_B(\eta) = \int_{-1}^1 d\cos\Theta ~p_\Omega(\cos\Theta) \delta[ \eta - \eta(\cos\Theta)]\,.
\end{equation}
Under a change of variables $\zeta = \eta(\cos\Theta)$, one finds
\begin{align}
	f_B(\eta) 
	 & = \frac{3}{4 \beta^* \cosh \eta^*}\int_{|\eta_\Upsilon - \eta^*|}^{\eta_\Upsilon + \eta^*} d\zeta ~ \frac{\delta[\eta - \zeta]\sinh{\zeta}}{\sinh\eta_\Upsilon}\bigg[1 - \bigg(\frac{\cosh\zeta - \cosh\eta^*\cosh\eta_\Upsilon}{\sinh\eta^*\sinh\eta_\Upsilon}\bigg)^2\bigg] \notag\\
	 & = \frac{3 \sinh \eta}{4 \sinh\eta^*\sinh\eta_\Upsilon}\bigg[1 - \bigg(\!\dfrac{\cosh\eta - \cosh\eta^*\cosh\eta_\Upsilon}{\sinh\eta^*\sinh\eta_\Upsilon}\!\bigg)^2\bigg]\,, \quad |\eta_\Upsilon - \eta^*| < \eta <  \eta^* + \eta_\Upsilon\,,\label{eqn:BRPDF}
\end{align} 
and zero otherwise. Note that $f_B$ itself has zeroes at each end of its non-trivial domain, i.e. at $\eta = \eta^* + \eta_\Upsilon$ and $|\eta_\Upsilon - \eta^*|$. The boost at the pdf peak is $\gamma_{\rm peak} = (\mathcal{E}_+ + \mathcal{E}_-)/(2m)$. E.g. for the Belle II parameters, the peak $\beta\gamma_{\rm peak} = \sinh \cosh^{-1}(\gamma_{\rm peak}) = 0.29$. This is the boost of $B$'s emitted at $\Theta = \pi/2$, and matches the quoted $B$ design boost at Belle II \cite{BelleII:2010td}.

\section{The Monte Carlo event generator}
\label{app:MC}

This appendix describes in more detail the MC event generator written in \texttt{C} and \texttt{Python}. The $B\to (K^*\to K\pi)\gamma$ phase space is generated as follows. The $B$ rapidity is sampled from the PDF given in eq.~\eqref{eqn:BRPDF} while the photon polar angle $\theta_\gamma$ and the $K$ polar and azimuthal angles are generated uniformly in the appropriate frame. On the other hand, since in BH photon conversion the leptons are produced with preferentially small angles with respect to the photon direction, the lepton polar angles are generated uniformly on a $\log$ scale. This is implemented via the transformed variables $t_\pm=\log_{10}\theta_\pm$ where $t_\pm$ are uniformly distributed and with $t_\pm\in[-5,-1]$. Moreover, the azimuthal angle separation between the leptons $(\delta\phi)$ is peaked around $\pi$ and so, to improve the efficiency of the generator, $\delta\phi$ is sampled from a Cauchy distribution. All other BH variables are generated uniformly.

The weight associated with each event is proportional to the matrix element~\eqref{eq:poldecsqamp}. The events are unweighted using the standard procedure. That is, the weights are normalized to the largest weight and the event is kept if its normalized weight is larger than a random number on $[0,1]$. Of course, this procedure assumes that the phase space was sufficiently sampled such that the largest weight found is close to the global maximum.

\begin{figure}[t!]
\includegraphics[width=0.495\columnwidth]{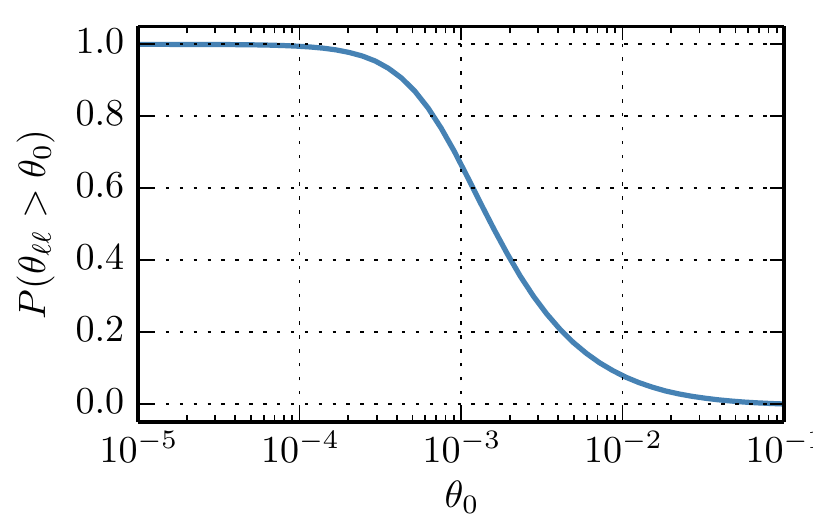}
\includegraphics[width=0.495\columnwidth]{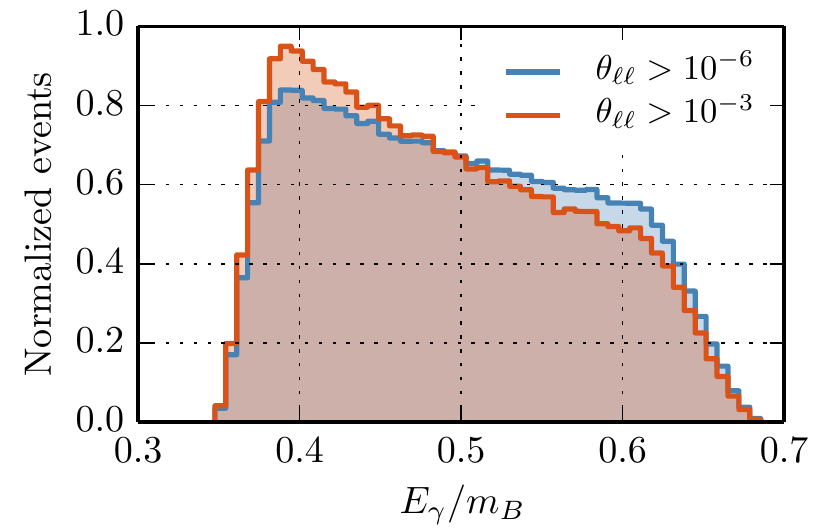}
\caption{Left: The cumulative distribution function (CDF) of the opening angle between the leptons. Right: The normalized distribution of the photon energy in units of the $B$ mass for two different $\theta_{\ell\ell}$ cuts.}
\label{fig:MC1}
\end{figure}

\begin{figure}[t!]
\includegraphics[width=0.495\columnwidth]{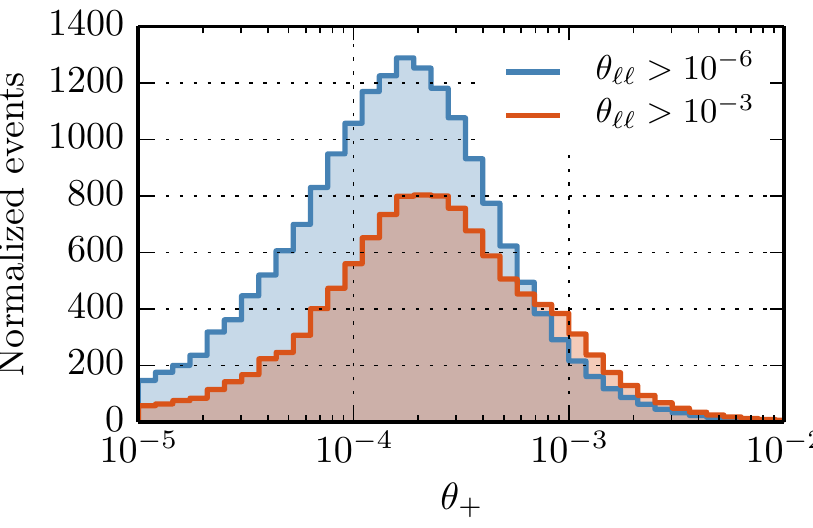}
\includegraphics[width=0.495\columnwidth]{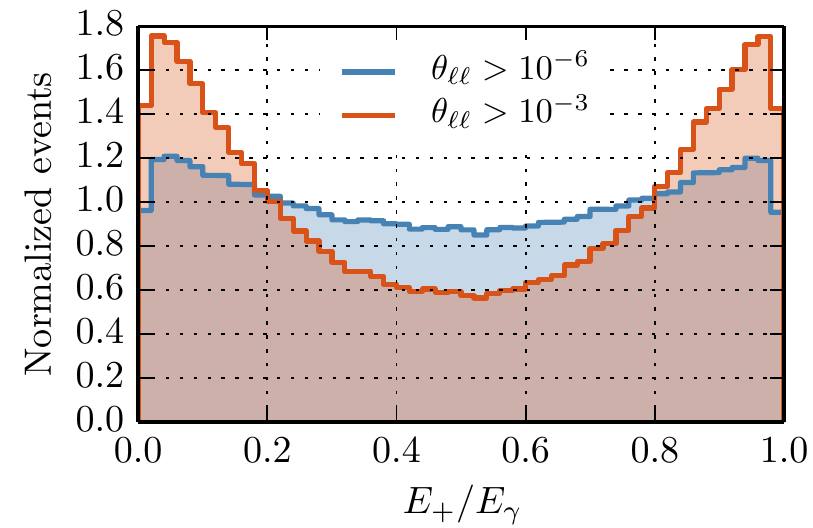}
\caption{Left: The normalized polar angle distribution of the positron for two different values of the opening angle cut $\theta_{\ell\ell}$. Right: the positron energy as a fraction of the photon energy for two values of $\theta_{\ell\ell}$. The distribution exhibits the expected behavior for BH conversion. It is symmetric about $1/2$ and prefers that one lepton carry a larger fraction of the photon energy.}
\label{fig:MC2}
\end{figure}

Using this procedure, we generate MC samples for many choices of $(r,\phi+\delta)$ couplets with 500k events per sample. Some representative distributions from the $(0.1,0)$ sample are shown in Figs~\ref{fig:MC1} and \ref{fig:MC2}. In particular, the left panel in Fig.~\ref{fig:MC1} shows the cumulative distribution function for the opening angle between the leptons $\theta_{\ell\ell}$ while the right panel shows the distribution of photons energies. Figure~\ref{fig:MC2} shows the polar angle and fractional energy distribution of the positron.

\FloatBarrier

\end{document}